\documentclass[twocolumn,aps,showpacs,superscriptaddress,floatfix]{revtex4}
\usepackage{amsmath}
\usepackage{graphicx}
\usepackage{dcolumn}
\usepackage{bm}
\usepackage{amssymb}
\usepackage{color}
\usepackage{xcolor}
\usepackage{xfrac}

\begin{document}

\title{Heat fluctuations and initial ensembles}

\author{Kwangmoo Kim}
\affiliation{Research Institute of Advanced Materials, Seoul National University, Seoul, 151-742, Korea}
\affiliation{School of Physics, Korea Institute for Advanced Study, Seoul, 130-722, Korea}

\author{Chulan Kwon}
\affiliation{Department of Physics, Myongji University, Yongin, Gyeonggi-do, 449-728, Korea}

\author{Hyunggyu Park}
\affiliation{School of Physics, Korea Institute for Advanced Study, Seoul, 130-722, Korea}

\date{\today}

\begin{abstract}

Time-integrated quantities such as work and heat increase incessantly in time during nonequilibrium processes near steady
states. In the long-time limit, the average values of work and heat become asymptotically equivalent to each other, since they only differ by a finite energy change in average.
However, the fluctuation theorem (FT) for the heat is found not to hold with the equilibrium initial ensemble,
while the FT for the work holds. This reveals an intriguing effect of everlasting initial
memory stored in rare events.
We revisit the problem of a Brownian particle in a harmonic potential dragged with a constant velocity,
which is in contact with a thermal reservoir. The heat and work fluctuations are investigated with initial Boltzmann
ensembles at temperatures generally different from the reservoir temperature.
We find that, in the infinite-time limit, the FT for the work is fully recovered
for arbitrary initial temperatures, while the heat fluctuations significantly deviate
from the FT characteristics except for the infinite initial-temperature limit
(a uniform initial ensemble).
Furthermore, we succeed in calculating finite-time corrections to the heat and work distributions analytically, using the modified saddle point integral method recently developed by us. Interestingly, we find non-commutativity between the infinite-time limit
and the infinite-initial-temperature limit for the probability distribution function (PDF) of the heat.

\end{abstract}


\maketitle

\section{\label{sec:level1}Introduction}

The fluctuation theorem (FT) has been regarded as a fundamental principle in nonequilibrium statistical mechanics. It concerns time-integrated quantities such as heat and work in nonequilibrium processes. The FT provides a rigorous rule for the thermal fluctuations of such quantities, independent of any detailed dynamics. The first form of the FT was found for entropy production piled in a heat bath for a deterministic thermostated system~\cite{evans93,evans94,gallavotti}, given as
$\langle e^{-\tau s}\rangle\to1$ as $\tau\to\infty$, where $\tau$  is the measuring time and $s$ is the entropy production rate in the unit of the Boltzmann constant $k_B$. Here the bracket $\langle\cdots \rangle$ denotes the average or integral over all the fluctuations. It is termed as a steady-state integral FT in literature.

Later, the FT was found to hold in stochastic systems~\cite{crooks,kurchan,lebowitz}. In particular, Crooks showed that the (transient) FT rigorously holds at all times (finite $\tau$) for the work $W$ produced in nonequilibrium systems starting from  equilibrium distributions. Furthermore, it can be expressed in a more detailed form, termed as a detailed FT, regarding the probability distribution function (PDF) of work fluctuations, given as
\begin{equation}
\frac{P_F(W)}{P_R(-W)}=e^{\beta (W-\Delta F)},
\label{eq:cft}
\end{equation}
where $\beta$ is the inverse temperature of the heat bath and $\Delta F$ the free energy difference between the initial and final times due to the change in a time-dependent protocol such as a volume, an external field, or a potential shape. $P_F$ denotes the PDF for the forward ($F$) process, while $P_R$ denotes that for the reverse ($R$) process where the protocol varies in time reversely to the forward process. The symmetry of the PDF such as
in Eq.\ (\ref{eq:cft}) is known, in general, as the Gallavotti-Cohen symmetry~\cite{gallavotti}.
The Jarzynski equality $\langle e^{-\beta W}\rangle=e^{-\beta \Delta F}$~\cite{jarzynski1} is nothing but the integral FT corresponding to the Crooks detailed FT. The discovery of the FT, which is expected to be valid in general stochastic systems, has resulted in extensive studies on unprecedented nonequilibrium phenomena. Many experimental evidences have also been reported~\cite{wang,carberry,douarche,gomez,ciliberto10}.

The choice of an initial ensemble is  critical for the validity of the FT. For example,
the transient detailed FTs for any finite $\tau$ hold and so do the integral FTs,
only with the equilibrium Boltzmann distribution as the initial ensemble for the work
or with the uniform (infinite-temperature) distributions for the heat~\cite{park,jslee1,noh}
(see also Sec.\ \ref{subsec:level33}).
The total entropy production satisfies the transient integral FT with an arbitrary initial ensemble~\cite{seifert}, though its detailed FT is valid only in the steady state.
In fact, the detailed FT guarantees the integral FT, but the converse is true only if the initial distributions for the forward and reverse paths are {\it involutary} to each other~\cite{esposito}.

A natural question arises as ``what happens to the FT when other types of
initial ensembles are taken?''. It is clear that the transient FT does not hold without
a proper initial ensemble corresponding to a time-integrated quantity, but, is
the (steady-state) FT in the $\tau\rightarrow\infty$ limit not valid for
the time-integrated quantities, either?
If not, how can the initial memory persist in the long-time limit?
In order to answer these questions, we investigate the effect of initial ensembles
on the detailed FT  for the heat and work, in particular for large $\tau$.

It is a formidable task to calculate the PDF exactly for finite $\tau$, so we restrict ourselves only to its large deviation function and corrections in the large $\tau$ limit. The PDF for a time-integrated quantity $A$ for a long period of time $\tau$ can be written in a scaling form
\begin{equation}
P(A)\sim e^{-\tau h(A/\tau)} \quad \text{for large}\ \tau,
\end{equation}
where $A$ is usually scaled dimensionless and
$h(A/\tau)$ is called a large deviation function (LDF).
Many interesting properties on the LDF were found on such as the current fluctuations~\cite{bodineaua,lacoste}, the (non-Gaussian) exponential tail~\cite{saito}, the everlasting initial memory threshold~\cite{jslee1,jslee2}, and so on. If the detailed FT  holds, the Gallavotti-Cohen (GC) symmetry is expressed in terms of the LDFs as
\begin{equation}
-h_F(A/\tau)+h_R(-A/\tau)=A/\tau ,
\label{eq:GC_LDF}
\end{equation}
where $h_F$ ($h_R$) is the LDF for the forward (reverse) process. When $A=\beta W$,
Eq.\ (\ref{eq:cft}) leads to the above symmetry in the large $\tau$ limit~\cite{exp1}.

The thermodynamic first law reads $\Delta E=W-Q$ for the energy change $\Delta E$. We define $Q$ as the heat transferred into the heat bath. In nonequilibrium close to the steady state, $\langle W\rangle$ and $\langle Q\rangle$ grow linearly in $\tau$, but $\langle\Delta E\rangle$ remains finite. Thus, one might expect that both quantities approximately have the identical PDFs for large $\tau$ as the difference $\Delta E$ may become negligible. Starting with the equilibrium Boltzmann ensemble, the detailed FT for $W$ is satisfied even at finite $\tau$, and thus is expected to be valid also for $Q$ at least in the infinite $\tau$ limit.
However, the reality is against the expectation. The detailed FT for the heat was examined
analytically for the motion of a particle in a harmonic potential dragged with a constant velocity, which is one of the experimental prototypes~\cite{wang,carberry,wang05,pesce}. It was found in the infinite $\tau$ limit that $\ln P_F(Q)/P_R(-Q)\simeq \beta Q$ only in the central region around $Q=0$~\cite{exp2}, while it approaches a plateau for large $|Q|$, which is the origin of the extended FT~\cite{vanzon1,vanzon2}. Recently, the modification of the detailed FT for the heat has been proposed for general systems in terms of correlations
between $\Delta E$ and $Q$~\cite{noh}.

The violation of the FT is due to a rare but non-negligible chance of $\Delta E$ having an extremely large value, which causes the FT modified in the tail region of the PDF~\cite{jslee1}.
The probability to find the initial system with an extremely large energy is exponentially small, but it will almost always dissipate the most of its energy into the reservoir in the long-time limit. Thus, this event becomes relevant to the tail part of the heat PDF which also decays exponentially for large $|Q|$. Even for very large $\tau$ and $|Q|$, there is
always an exponentially small probability to find the event with the corresponding large energy in the initial Boltzmann ensemble. Therefore, this effect can not go away even in the long-time limit. This so-called ``boundary effect'' recognized in many references~\cite{farago,visco,puglisi,sabhapandit} is  observed for an unbounded energy distribution in the initial ensemble, but obviously not observed when the initial energy is bounded.

As the initial ensemble plays a crucial role in the FT violation,
we study its effect on the work and heat fluctuations more systematically
in this paper. As an initial ensemble, we take the Boltzmann distribution at a temperature generally different from that of the heat reservoir. In this case, the FTs for both $W$ and $Q$  do not hold for finite $\tau$. However, it is not obvious whether
the FT will hold or not in the large $\tau$ limit. The validity may depend on the quantity of interest and also on the temperature of the initial ensemble.
In fact, it is already
reported that the injected and dissipated PDF's of heat in an equilibration process show phase transitions at two different finite initial temperatures, respectively,
below which the LDF is not affected, while
above which the LDF is significantly modified by the boundary term~\cite{farago,jslee1}.

In this paper, we revisit the problem of a Brownian particle in a harmonic potential dragged with a constant velocity, which is in contact with the thermal reservoir.
We then investigate the PDFs of the work and heat for a long period of $\tau$. For the heat PDF,
we find the singularities due to the boundary terms, which vary with
the temperature of the initial ensemble and break the GC symmetry of the PDF.
As the initial temperature approaches the infinity in the infinite-$\tau$ limit,
the GC symmetry is restored.
We also calculated a finite-$\tau$ correction for the heat PDF, where the singularity
structure becomes more complicated. Using the modified saddle point integral method recently
developed by us~\cite{jslee2}, we exactly obtained the LDF of the heat up to
${\cal O}(\tau^{-1})$ and thus the FT violation is measured up to the same order.
Interestingly, the finite-$\tau$ correction of the FT violation does not vanish in the infinite initial-temperature limit, which implies the non-commutativity between the two limits of the infinite $\tau$ and the infinite initial temperature. However, we can show that the
transient FT is satisfied for any $\tau$ if one takes a proper infinite initial-temperature limit
before taking the infinite-$\tau$ limit.

In contrast, the work PDF turns out to be free of any singularity even at any initial temperature. Furthermore, the work PDF can be calculated exactly
at any finite $\tau$ and any initial temperature. We can show that
the transient FT does not hold except when the initial temperature is identical to the temperature of the reservoir. However, in the infinite-$\tau$ limit, the FT is fully restored, regardless of the initial temperature. The difference between
the FT violations for the heat and work comes from the presence of $\Delta E$, which
induces everlasting initial memory in the heat PDF.

The remainder of this paper is organized as follows. In Sec.\ \ref{sec:level2},
we introduce a model and theoretical formalism to obtain the PDF of the heat and work.
The generating functions for the heat and work PDF are derived.
In Sec.\ \ref{sec:level3}, we present the LDF and the FT violation for the heat
fluctuations in the long-time limit and their finite-time corrections.
The restoration of the FT for the heat in the infinite initial-temperature limit is also discussed.
In Sec.\ \ref{sec:level4}, we repeat the calculations for the work fluctuations.
Finally, in Sec.\ \ref{sec:level6}, we summarize our study and discuss the physical origin of the everlasting initial memory in the time-accumulated quantities. In Appendix, the exact
generating functions for the heat and work are given at finite $\tau$.

\section{\label{sec:level2}Model and Generating functions}

\subsection{\label{subsec:level21}Model}

The Brownian motion of a particle in a moving harmonic potential with a constant velocity $\mathbf{v}^\ast$~\cite{vanzon1},
is described by an overdamped Langevin equation as
\begin{equation}
\dot{\mathbf{x}}_t=-\tau_r^{-1}(\mathbf{x}_t-\mathbf{x}_t^\ast)+\alpha^{-1}\bm{\zeta}_t,
\label{eq:ole}
\end{equation}
where $\mathbf{x}_t$ is the position of the particle at time $t$, $\tau_r$ the relaxation time, $\mathbf{x}_t^\ast=\mathbf{v}^\ast t$ the moving center of
the harmonic potential, and $\alpha$ the Stokes friction of the particle in a fluid.
The relaxation time is given by $\tau_r=\alpha/k$, where $k$ is the force constant of the harmonic potential. $\bm{\zeta}_t$ is a fluctuating white noise given as
\begin{equation}
\langle \zeta_{t}^a\rangle=0,\quad \langle \zeta_{t}^a\zeta_{t^\prime}^{a^\prime} \rangle=2\alpha k_BT\delta_{aa^\prime}\delta(t-t^\prime)
\label{eq:zeta}
\end{equation}
where the superscript $a$ and $a^\prime$ denote component indices ($a,a^\prime=1,\cdots,d$) for a $d$-dimensional motion and $T$ is the temperature of the heat bath. The particle and the center of the harmonic potential are initially positioned
at the origin: $\mathbf{x}_0=\mathbf{x}_0^\ast=0$.

For convenience, we first find out the {\em deterministic} part $\mathbf{y}_t^\ast$ of the solution to Eq.\ (\ref{eq:ole}) as
\begin{equation}
{\mathbf{y}}_t^\ast=\mathbf{v}^\ast t-\mathbf{v}^\ast\tau_r(1-e^{-t/\tau_r}),
\label{eq:yt}
\end{equation}
satisfying the deterministic equation
$\dot{\mathbf{y}}_t^\ast=-\tau_r^{-1}(\mathbf{y}_t^\ast-\mathbf{x}_t^\ast)$
with an initial condition $\mathbf{y}_0^\ast=0$.  If we look at the relative
motion of the particle as
\begin{equation}
\mathbf{X}_t=\mathbf{x}_t-\mathbf{y}_t^\ast,
\label{eq:devavgm}
\end{equation}
then it satisfies a simpler equation of motion as
\begin{equation}
\dot{\mathbf{X}}_t=-\tau_r^{-1}\mathbf{X}_t+\alpha^{-1}\bm{\zeta}_t.
\label{eq:simole}
\end{equation}

The harmonic potential energy $U_t=\frac{k}{2}(\mathbf{x}_t - \mathbf{x}_t^\ast)^2$ has an explicit time dependence. As recognized by Jarzynski~\cite{jarzynski1}, the work is transferred into the system by the rate of $\partial U_t/\partial t$. It is performed by an external agent (experimental device) to change the protocol $\mathbf{x}_t^\ast$. Then, the work $W_\tau$ delivered into the system can be expressed along a given
trajectory $[\mathbf{x}_t]_0^\tau$ for $0\le t \le\tau$ as
\begin{eqnarray}
W_\tau & = & -k\int_0^\tau \mathrm{d}t (\mathbf{x}_t-\mathbf{x}_t^\ast)\cdot \mathbf{v}^\ast\nonumber \\
& = & -k \int_0^\tau \mathrm{d}t[\mathbf{v}^\ast\cdot\mathbf{X}_t+\mathbf{v}^\ast\cdot(\mathbf{y}_t^\ast
-\mathbf{x}_t^\ast)].
\label{eq:w}
\end{eqnarray}
The heat $Q_{\tau}$ going into the fluid along the same trajectory $[\mathbf{x}_t]_0^\tau$ is given by
\begin{equation}
Q_{\tau}= W_{\tau}-\Delta U_{\tau},
\label{eq:q}
\end{equation}
where $\Delta U_{\tau}= U_{\tau}-U_0$ is the potential energy change. Note that only the potential energy change is considered in the overdamped limit.

The PDF for the work or heat can be obtained by considering all the possible trajectories.
For convenience, we scale
the heat and work by the temperature of the heat bath to get dimensionless quantities as
$\beta Q$ and $\beta W$ with $\beta=1/(k_B T)$. The finite-$\tau$ PDF for a quantity $A_\tau$
($=\beta Q_\tau$ or $\beta W_\tau$) can
be written as
\begin{eqnarray}
\label{eq:pdf}
P_\tau (A) &=& \left\langle \delta (A-A_\tau)\right\rangle \nonumber\\
&=& \int_{-i\infty}^{+i\infty}\ \frac{\mathrm{d}\lambda}{2\pi i} \ e^{\lambda A}
\left\langle e^{-\lambda A_\tau} \right\rangle,
\end{eqnarray}
where $A_\tau$ is the trajectory-dependent fluctuating quantity  and
$\langle \cdots\rangle$ denotes an average over all the possible trajectories and the initial distribution.

It is convenient to introduce a generating function defined as
\begin{equation}
g_A(\lambda)\equiv \left\langle e^{-\lambda A_\tau} \right\rangle=\int_{-\infty}^{+\infty} \mathrm{d}A \ P_\tau (A)\ e^{-\lambda A}.
\label{eq:generating_function}
\end{equation}
Then, the PDF is simply a Fourier transform of the generating function as in Eq.\ (\ref{eq:pdf}). The GS symmetry in terms of the generating function can be obtained,
using Eq.\ (\ref{eq:GC_LDF}), as
\begin{equation}
g_{A}(\lambda)=g_{A} (1-\lambda),
\label{eq:GC_g}
\end{equation}
where the process indices, $F$ and $R$, are dropped because the generating
functions for the forward and reverse processes are equivalent to each other in our constantly
moving harmonic potential. Any energetic quantity like heat or work does not
depend on the sign of the velocity $\mathbf{v}^\ast$ of the moving harmonic potential. Furthermore,
the free energy difference in Eq.\ (\ref{eq:cft}) is always zero ($\Delta F=0$),
since the shape of the potential energy does not change except for a translation.

In order to study the influence of an initial condition, we assume that the particle initially has an equilibrium distribution at the initial inverse temperature $\beta^\prime$ as
\begin{equation}
\rho_{\mathrm{in}}(\mathbf{X}_0)=\left(\frac{\beta^\prime k}{2\pi}\right)^{d/2}e^{- \frac{\beta^\prime k}{2}\mathbf{X}_0^2}.
\label{eq:initiald}
\end{equation}

\subsection{\label{subsec:level22}Generating function for heat}

The generating function $g_Q (\lambda)$ for the heat is written as
\begin{eqnarray}
g_Q(\lambda) & = & \left\langle e^{-\lambda\beta Q_\tau}\right\rangle=\left\langle e^{-\lambda\beta(W_\tau-\Delta U_\tau)}\right\rangle \nonumber \\
& = & \int \mathrm{d}\mathbf{X}_{\tau}e^{\frac{\lambda\beta k}{2}(\mathbf{X}_{\tau}+\Delta\mathbf{y}_{\tau}^\ast)^2}\int \mathrm{d}\mathbf{X}_0 \rho_{\mathrm{in}}(\mathbf{X}_0)e^{-\frac{\lambda\beta k}{2}\mathbf{X}_0^2} \nonumber \\
&\times &  \int D[\mathbf{X}_{t}]e^{\int_0^\tau \mathrm{d}t{\mathcal{L}}}e^{\lambda\beta k\int_0^\tau \mathrm{d}t(\mathbf{v}^\ast\cdot \mathbf{X}_t+\mathbf{v}^\ast\cdot\Delta\mathbf{y}_t^\ast)},
\label{eq:g1}
\end{eqnarray}
where $\Delta\mathbf{y}_t^\ast=\mathbf{y}_t^\ast-\mathbf{x}_t^\ast=-\mathbf{v}^\ast
\tau_r(1-e^{-t/\tau_r})$ from Eq.\ (\ref{eq:yt}) and $\int D[\mathbf{X}_t]$ denotes the path integral over all the trajectories connecting $\mathbf{X}_0$ and $\mathbf{X}_\tau$,
with proper normalizations. The Lagrangian $\mathcal{L}$ is given in a pre-point (Ito) representation for the time discretization~\cite{discret} as
\begin{equation}
{\mathcal{L}}=-\frac{1}{4D}\left(\dot{\mathbf{X}}_t+\tau_r^{-1}\mathbf{X}_t\right)^2
\end{equation}
for $D=(\beta \alpha)^{-1}$.

Noting that $\mathcal{L}$ is quadratic in $\mathbf{X}_t$, the generating function is in fact a succession of a multivariate Gaussian integral over $\mathbf{X}_{j}$ at discretized times
$t_{j}=j \tau/N$ ($j=0,1,\cdots, N$) with a large $N$. We can compute the integral in the $N\to\infty$ limit by using the method in our previous work~\cite{kwon}.
It is convenient to rewrite the generating function in terms of
normalized Gaussian integrations over $\{\mathbf{X}_{j}\}$ $(0\le j\le N)$ as
\begin{equation}
g_Q(\lambda)=c \mathcal{N}\left\langle
e^{\lambda\beta k(\Delta\mathbf{y}_{\tau}^{\ast}\cdot\mathbf{X}_{\tau}+\mathbf{v}^{\ast}
\cdot\int_0^\tau\mathrm{d}t\mathbf{X}_t)}\right\rangle_{\{\mathbf{X}_j\}},
\end{equation}
where the average is defined as
\begin{eqnarray}
\langle {\cal O} \rangle_{\{\mathbf{X}_j\}} & = & \frac{1}{\mathcal{N}}\int \mathrm{d}\mathbf{X}_{N}e^{\frac{\lambda\beta k}{2}\mathbf{X}_{N}^2}\int \mathrm{d}\mathbf{X}_0\rho_{\mathrm{in}}(\mathbf{X}_0)e^{-\frac{\lambda\beta k}{2}\mathbf{X}_0^2} \nonumber \\
& & \times\int \prod_j \mathrm{d}\mathbf{X}_j e^{\int_0^\tau \mathrm{d}t\mathcal{L}}\ {\cal O},
\label{eq:norm_factor}
\end{eqnarray}
with the normalization constant $\mathcal{N}$ obtained from
$\langle 1 \rangle_{\{\mathbf{X}_j\}}=1$.
The non-fluctuating deterministic part yields
\begin{equation}
c=e^{\frac{\lambda\beta k}{2}(\Delta \mathbf{y}_\tau^{\ast})^2+\lambda\beta k\mathbf{v}^{\ast}\cdot\int_0^{\tau} \mathrm{d}t\Delta \mathbf{y}_t^{\ast}}.
\end{equation}

As the distribution in the above average is a simple Gaussian, it is sufficient to consider the cumulants only up to the second order. It is then straightforward to find
\begin{eqnarray}
g_Q(\lambda)&=&c\mathcal{N} e^{\frac{(\lambda\beta k)^2}{2}|\mathbf{v}^\ast|^2\int_0^\tau \mathrm{d}t\int_0^\tau \mathrm{d}t^\prime C(t,t^\prime)} \nonumber \\
& \times & e^{\frac{(\lambda\beta k)^2}{2}\left[|\Delta\mathbf{y}_{\tau}^\ast|^2C(\tau,\tau)
+2\Delta\mathbf{y}_{\tau}^\ast\cdot\mathbf{v}^\ast\int_0^{\tau}\mathrm{d}t C(\tau,t)\right]},
\label{eq:g_Q}
\end{eqnarray}
where $C(t,t^\prime)$ is a correlation function given by
\begin{equation}
\langle X_{t}^a X_{t^\prime}^{a^\prime}\rangle_{\{\mathbf{X}_i\}}=\delta_{aa^\prime} C(t,t^\prime).
\end{equation}

The integrations at the initial and final points in Eq.\ (\ref{eq:norm_factor}) include the extra boundary factors $e^{-\lambda\beta k\mathbf{X}_0^2/2}$ and $e^{\lambda\beta k\mathbf{X}_{N}^2/2}$, respectively, which modify
$\mathcal{N}$ and $C(t,t')$ significantly. After some algebra with
the initial Boltzmann condition with the inverse temperature $\beta^\prime$ in Eq.\ (\ref{eq:initiald}), we find
\begin{equation}
\mathcal{N}=\left(\frac{\beta^\prime}{\beta^\prime+\lambda \beta}\right)^{d/2}
\left(\frac{1}{1-\lambda\beta kA_\tau^{-1}}\right)^{d/2}
\label{eq:N}
\end{equation}
and
\begin{equation}
C(t,t^\prime) = e^{-\frac{t-t^\prime}{\tau_r}}A_{t^\prime}^{-1}\frac{1-\lambda\beta k a_{\tau-t}^{-1}}{1-\lambda\beta kA_\tau^{-1}} \quad \text{for}\ t\ge t^\prime,
\label{eq:cumul}
\end{equation}
where
\begin{eqnarray}
A_t^{-1} & = & (\beta k)^{-1}(1-e^{-\frac{2t}{\tau_r}}) \nonumber \\
& & +(\beta^\prime k+\lambda\beta k)^{-1}e^{-\frac{2t}{\tau_r}} \nonumber \\
a_{\tau-t}^{-1} & = & (\beta k)^{-1}(1-e^{-\frac{2(\tau-t)}{\tau_r}})~.
\end{eqnarray}
Note that $A_t$ is the Gaussian kernel at time $t$ during the path integral.
Without any extra term, $\mathcal{N}=1$ and $C(t,t^\prime)=e^{-\frac{t-t^\prime}{\tau_r}} A_{t^\prime}^{-1}$ with $A_{t^\prime}^{-1}=(\beta k)^{-1}$.

For simplicity, we adopt the same parameter values and notations in Ref.\ \cite{vanzon2} as
\begin{equation}
\beta=1, \quad k=1, \quad \tau_r=1, \quad \text{and}\quad |\mathbf{v}^\ast|^2=w.
\end{equation}
In these units, $w$ is equal to the average work rate in the steady state: $w=\lim_{\tau\to\infty}\langle W_\tau\rangle/\tau$. Putting all together into Eq.\ (\ref{eq:g_Q}), we find the exact expression for $g_Q(\lambda)$, which is
quite complicated and
shown in Appendix~\ref{sec:app1}. In the following, we will evaluate the LDF up to the order of $\tau^{-1}$, so here we ignore all the exponentially decaying terms like $e^{-\tau}$ in
$g_Q(\lambda)$. Then,
we get a rather simple form as
\begin{equation}
g_Q(\lambda)=
\frac{\beta^{\prime d/2} e^{ - w\left[\tau\lambda(1-\lambda) - \frac{3}{2}\lambda +\frac{\lambda^2}{2}\left(4-\frac{1}{\beta^\prime+\lambda}\right)\right]}}
{\left[(\beta^\prime+\lambda)(1-\lambda)\right]^{d/2}}.
\label{eq:g3}
\end{equation}
Note that the GC symmetry in Eq.\ (\ref{eq:GC_g}) seems to be  preserved
at the level of the large deviation function (exponent) in the $\tau=\infty$ limit. However, the singular property of the prefactor coming from the boundary terms does not
uphold the GC symmetry, which causes a significant violation of the GC symmetry
in the heat PDF, even in the $\tau=\infty$ limit.

\subsection{\label{subsec:level23}Generating function for work}

The generating function $g_W(\lambda)$ for the work is given as
\begin{eqnarray}
g_W(\lambda) &=&\langle e^{-\beta\lambda W_\tau}\rangle \nonumber \\
& = & \int \mathrm{d}\mathbf{X}_\tau\int \mathrm{d}\mathbf{X}_0 \rho_{\mathrm{in}}(\mathbf{X}_0) \int D[\mathbf{X}_t]\ e^{\int_0^\tau \mathrm{d}t\mathcal{L}} \nonumber \\
&\times& e^{\lambda\beta k\int_0^\tau \mathrm{d}t(\mathbf{v}^\ast\cdot\mathbf{X}_t
+\mathbf{v}^\ast\cdot\Delta\mathbf{y}_t^\ast)}.
\label{eq:g_work}
\end{eqnarray}
Similarly, we get
\begin{eqnarray}
g_W(\lambda)&=&e^{\lambda\beta k\mathbf{v}^{\ast}\cdot\int_0^{\tau} \mathrm{d}t\Delta \mathbf{y}_t^{\ast}}e^{\frac{(\lambda\beta k)^2}{2}|\mathbf{v}^\ast|^2} \nonumber \\
&\times& e^{\int_0^\tau \mathrm{d}t\int_0^\tau \mathrm{d}t^\prime C(t,t^\prime)},
\end{eqnarray}
with the correlation function
\begin{equation}
C(t,t^\prime)= e^{-\frac{t-t^\prime}{\tau_r}}A_{t^\prime}^{-1} \quad \text{for}\ t>t^\prime,
\label{eq:cumulw}
\end{equation}
where
\begin{equation}
A_t^{-1} = (\beta k)^{-1}(1-e^{-\frac{2t}{\tau_r}})+(\beta^\prime k)^{-1}e^{-\frac{2t}{\tau_r}}.
\end{equation}

Using the same convention ($\beta=1$, $k=1$, $\tau_r=1$, and $|\mathbf{v}^\ast|^2=w$) and neglecting the exponentially decaying terms like $e^{-\tau}$ (see the full solution in Appendix \ref{sec:app1}), we find
\begin{equation}
g_W(\lambda)
= e^{-w\left[ \tau\lambda(1-\lambda)-\lambda+\frac{\lambda^2}{2}\left(3-\frac{1}{\beta^\prime}\right)\right]}.
\label{eq:c}
\end{equation}
The GC symmetry is satisfied only in the $\tau=\infty$ limit, but for an arbitrary $\beta^\prime$. At $\beta^\prime=1$, it holds for any finite $\tau$ as expected, even when the exponentially decaying terms are included in $g_W(\lambda)$ (see Appendix \ref{sec:app1}).

\section{\label{sec:level3}LDF and FT for heat}

\subsection{\label{subsec:level31}Long-time limit}

As $\tau\to \infty$ in Eq.\ (\ref{eq:g3}), the generating function $g_Q(\lambda)$ exhibits the large deviation behavior as
\begin{equation}
g_Q(\lambda) \sim e^{-w\tau e(\lambda)}
\end{equation}
with
\begin{equation}
e(\lambda)=\left\{\begin{array}{ll}\lambda(1-\lambda) & \ \mbox{for $-\beta^\prime<\lambda<1$} \\ -\infty & \ \mbox{otherwise,}\end{array}\right.
\end{equation}
where the divergence is evident as $\lambda \rightarrow 1$ from below
and $-\beta^{\prime}$ from above. Each of them is due to the boundary term
at the final and initial points, respectively.

As the PDF $P_\tau (Q)$ is given by the Fourier transformation of $g_Q(\lambda)$
as in Eq.\ (\ref{eq:pdf}), we expect for large $\tau$
\begin{equation}
P_\tau(Q)\sim e^{-w\tau h(p)} \quad \text{with} \quad  p\equiv Q/(w\tau),
\label{eq:P_Q_p}
\end{equation}
where $p$ is a properly scaled variable for the heat.
Then, the LDF $h(p)$ is simply given by
the Legendre transform of $e(\lambda)$, given by
\begin{equation}
h(p)=\max_\lambda[e(\lambda)-\lambda p].
\label{eq:epdef}
\end{equation}
We find
\begin{equation}
h(p)=\left\{\begin{array}{ll}-p & \text{for}\ p<-1 \\ (p-1)^2/4 &
\text{for}\ -1\leq p\leq 2\beta^\prime+1 \\ \beta^\prime p-\beta^\prime(1+\beta^\prime) & \text{for}\ p>2\beta^\prime+1. \end{array}\right.
\label{eq:LDF}
\end{equation}
Note that the non-analytic behavior of the LDF $h(p)$ originates from
the divergence of $e(\lambda)$ due to the prefactor singularity in $g_Q(\lambda)$.

The detailed FT for the heat is examined by
\begin{equation}
f_\tau(p)=\frac{1}{\omega\tau}\ln\left[\frac{P_\tau(w\tau p)}{P_\tau(-w\tau p)}\right],
\label{eq:fq}
\end{equation}
where $f_\tau(-p)=-f_\tau(p)$. If the transient detailed FT is satisfied, then $f_\tau (p)=p$ for any $\tau$.
In the $\tau\to \infty$ limit, we can easily find $f_\infty (p)=-h(p) + h(-p)$, yielding
\begin{equation}
f_{\infty}(p)=\left\{\begin{array}{ll} p & \ \text{for}\ 0\leq p<1 \\ p-(p-1)^2/4 & \ \text{for}\ 1\leq p<2\beta^\prime+1 \\ (1-\beta^\prime)p+\beta^\prime(1+\beta^\prime) & \ \text{for}\ p\geq 2\beta^\prime+1.\end{array}\right.
\label{eq:ssft}
\end{equation}

\begin{figure}[ht!]
\begin{center}
\includegraphics[width=\columnwidth]{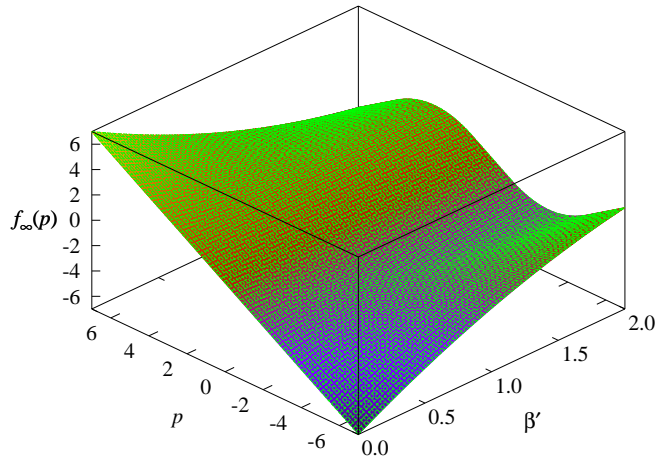}\\%
\end{center}
\caption{\label{fig:betapphase01}(Color online) $f_{\infty}(p)$ as in Eq.\ (\ref{eq:ssft}) is plotted as a function of $p$ and $\beta^\prime$.}
\end{figure}

\begin{figure}[ht!]
\begin{center}
\includegraphics[width=\columnwidth]{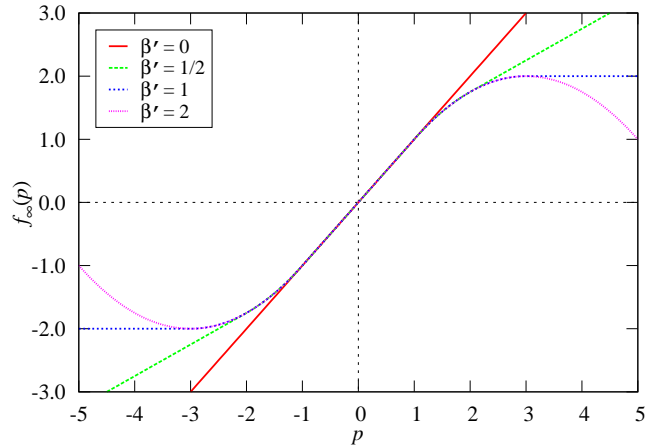}\\%
\end{center}
\caption{\label{fig:fqbetap01}(Color online) $f_{\infty}(p)$ for $\beta^\prime=0$, $1/2$, $1$, and $2$.}
\end{figure}

Indeed, the detailed FT for the heat holds only for $ |p|\le 1$ (region I), outside of which $f_{\infty}(p)$ deviates significantly from the FT relation. Its deviation depends on the initial temperature $\beta^\prime$ and differs in $1\leq |p|<2\beta^\prime+1$ (region II) and in $|p|\geq 2\beta^\prime+1$ (region III), as seen in Figs.\ \ref{fig:betapphase01} and \ref{fig:fqbetap01}. It is interesting to note that the FT is restored for all $p$ in the $\beta^\prime=0$ (infinite initial-temperature) limit, where the region II disappears and  $f_\infty (p)$ approaches $p$ in the region III. We will be back to this limit later in this section. The extended FT discussed by van Zon and Cohen~\cite{vanzon1,vanzon2} is a special case at $\beta^\prime=\beta=1$.

\subsection{\label{subsec:level32}Finite-time corrections}

It is difficult to compare the results in the $\tau\to\infty$ limit with those in the simulations or experiments, due to huge sampling  errors in the PDF tail representing rare events. In particular, the FT violation appears in this tail region. It is thus desirable to estimate finite-time corrections analytically. We want to evaluate the LDF up to $\mathcal{O}(\tau^{-1})$.

From Eq.\ (\ref{eq:pdf}), the PDF for the heat $Q$ $(=w\tau p)$ is written as
\begin{equation}
P_{\tau}(w\tau p) = \int_{-i\infty}^{+i\infty} \mathrm{d}\lambda~ \phi(\lambda)~ e^{-w\tau H(\lambda,p)},
\label{eq:FourierPDF}
\end{equation}
with the prefactor
\begin{eqnarray}
\phi(\lambda) & = & \frac{{\beta^\prime}^{d/2}}{2\pi i\left[(\beta^\prime+\lambda)(1-\lambda)\right]^{d/2}} \nonumber \\
& \times & e^{w\left[\frac{3\lambda}{2}-2\lambda^2+\frac{\lambda^2}{2(\beta^\prime +\lambda)}\right]},
\end{eqnarray}
and
\begin{equation}
H(\lambda,p) = -\lambda p +\lambda (1-\lambda).
\label{eq:el}
\end{equation}
The prefactor shows singularities at $\lambda=\lambda_1=1$ and $\lambda_2=-\beta^\prime$, which are simple poles for $d=2$, but branch points for $d=1, 3$.
Later, we choose a branch cut on the real axis of $\lambda$ for $\lambda>\lambda_1$ and $\lambda<\lambda_2$ when $d\neq 2$. In addition, there is an essential singularity at $\lambda=\lambda_2$, which will cause a little more complication in the following integration.

The integral for large $\tau$ can be approximated by the integral along the steepest descent path passing through a saddle point in the complex plane of $\lambda$. In the conventional saddle-point approximation, a saddle point $\lambda_0^*(p)$ is chosen by extremizing $H(\lambda,p)$ such as $\mathrm{d}H/\mathrm{d}\lambda|_{\lambda_0^*}=0$, yielding
$\lambda_0^*=(1-p)/2$. However, the integral may diverge due to the prefactor $\phi(\lambda)$ when the saddle point approaches one of its singularities.

In this study, we adopt the modified saddle point integral method~\cite{jslee2}
and search for the modified saddle points $\lambda^*(p)$ by extremizing
\begin{equation}
S(\lambda,p)\equiv H(\lambda,p)-(w\tau)^{-1} \ln \phi (\lambda),
\end{equation}
with
\begin{equation}
\left.\frac{\mathrm{d} S}{\mathrm{d} \lambda}\right|_{\lambda=\lambda^*}=0.
\label{eq:elp}
\end{equation}
There are multiple saddle points for a given $p$. However, it can be shown that there always exists a saddle point $\lambda^*(p)$ on the real-$\lambda$ axis between the two singularities, i.e., $\lambda_2<\lambda^{\ast}(p)<\lambda_1$. This saddle point is
$\tau$-dependent and sometimes approaches the singularities asymptotically
for large $\tau$. For $-1< p < 2\beta^\prime +1$, $\lambda^*$ approaches
the conventional saddle point $\lambda_0^*$, otherwise one of the singularities such as
$\lambda_1$ for $p<-1$ and $\lambda_2$ for $p>2\beta^\prime +1$, respectively.

When the modified saddle point $\lambda^*$ is nearby the singularities, the integral
along the steepest descent path should be performed with special care, because it
becomes a non-Gaussian integral, described in detail in the Appendix of Ref.\ \cite{jslee2}.

Now we present the results for different regions of $p$ as follows.

\subsubsection{\label{subsubsec:level321}The central region of the PDF}

Sufficiently deep inside of the interval of $-1<p<2\beta^\prime +1$, the saddle point $\lambda^\ast$ is given by
\begin{equation}
\lambda^\ast = \frac{1-p}{2}+\mathcal{O}(\tau^{-1}),
\label{eq:central}
\end{equation}
which approaches $\lambda_0^*$ for large $\tau$ and is far enough from the singularities at $\lambda_1$ and $\lambda_2$. Thus, one can apply the conventional saddle point approximation (see Eq.\ (A.20) in Ref.\ \cite{jslee2}),
which yields
\begin{eqnarray}
P_\tau(w\tau p)&=&i\left[\frac{2\pi}{w\tau |H^{\prime\prime}(\lambda_0^*)|}\right]^{1/2}
\phi(\lambda_0^*) e^{-w\tau H(\lambda_0^*)} \nonumber \\
&=&\sqrt{\frac{\beta^{\prime d}}{\pi w\tau}}\frac{2^{d-1}}{\left[(1+p)(2\beta^\prime+1-p)\right]^{d/2}} \nonumber \\
& \times & \ e^{-w\tau \frac{(1-p)^2}{4} +\frac{3}{4}w(1-p)} \nonumber \\
&\times & e^{-\frac{w(1-p)^2\left[2(1-p)+4\beta^\prime-1\right]}{4(2\beta^\prime+1-p)}}.
\label{eq:PDF_central}
\end{eqnarray}
This result is exact up to $\mathcal{O}(\tau^{-1})$ for the $\tau$-dependent LDF defined as
\begin{eqnarray}
h_\tau(p)&\equiv& -\frac{1}{w\tau}\ln P_\tau(w\tau p) \nonumber \\
&=& h_c(p) +\frac{\ln \tau }{2w\tau} -\frac{r_c(p)}{w\tau},
\label{eq:tLDFc}
\end{eqnarray}
where $h_c(p)=(1-p)^2/4$ and $r_c(p)$ is the logarithm of the ${\cal O}(1)$ terms in the
multiplicative factor and also in the exponent in Eq.\ (\ref{eq:PDF_central}).
The usual asymptotic LDF in Eq.\ (\ref{eq:LDF})
is obtained as $h(p)=\lim_{\tau\rightarrow\infty} h_\tau(p)$.

\subsubsection{\label{subsubsec:level322}The left wing of the PDF}

The saddle point $\lambda^*$ approaches the singularity at $\lambda=\lambda_1$ $(=1)$ from below,
in the left side of the central region ($p\lesssim -1$). Let us write
$\delta\lambda_{1}=\lambda^*-\lambda_1$ $(<0)$. For small $\delta\lambda_1$ and large $\tau$, the saddle-point equation (\ref{eq:elp}) is expanded in $\delta\lambda_1$ as
\begin{equation}
-2\delta\lambda_1-(p+1)+\frac{d}{2w\tau\delta\lambda_1}\approx 0.
\label{eq:elpmi}
\end{equation}
Its proper solution is
\begin{equation}
\delta\lambda_1=\frac{1}{4}\left[-(p+1)-\sqrt{(p+1)^2+\frac{4d}{w\tau}}\right].
\label{eq:dl}
\end{equation}

For $(p+1)\tau^{1/2}\ll -1$, Eq.\ (\ref{eq:dl}) becomes
\begin{equation}
\delta\lambda_1 \approx\frac{d}{2w(p+1)}\tau^{-1},
\label{eq:asympt_left}
\end{equation}
which determines the PDF in the most region of $p<-1$.
Note that the saddle point is already very close to $\lambda_1$ with a distance of
$\mathcal{O}(\tau^{-1})$.

For $|p+1|\tau^{1/2}\ll 1$, Eq.\ (\ref{eq:dl}) becomes
\begin{equation}
\delta\lambda_1\approx -\frac{1}{2}\sqrt{\frac{d}{w}}\tau^{-1/2},
\label{eq:asympt_left_boundary}
\end{equation}
which determines the PDF in a narrow region around $p=-1$.
This region vanishes in the $\tau=\infty$ limit.
In this case, the distance between the saddle point and $\lambda_1$ shrinks slower
with a distance of $\mathcal{O}(\tau^{-1/2})$.

The steepest descent integration passing through the saddle point near the singularity
becomes problematic, mainly because the singular prefactor cannot be expanded around
the singularity. However, the integration can be still performed only with
the expansion of the exponent $H(\lambda,p)$ around the saddle point.
The price to pay is that one should perform a non-Gaussian integration along the steepest
descent path. The integration results are explicitly given in the Appendix of Ref.\ \cite{jslee2} for general power-law singularities. Here, we just briefly sketch
the integration method.

We expand $H(\lambda,p)$ in powers of $\delta \lambda_1$  and  use a new variable $v$ defined as $v=1+(\lambda-\lambda^*)/\delta\lambda_1$. Then, Eq.\ (\ref{eq:FourierPDF}) can be written as
\begin{equation}
P_\tau(w\tau p) = C_1\int_{1-i\infty}^{1+i\infty}\mathrm{d}v\frac{e^{w\tau\left[(p+1)\delta\lambda_1 v+
\delta\lambda_1^2 v^2\right]}}{v^{d/2}},
\label{eq:PDF_left}
\end{equation}
where
\begin{equation}
C_1=\frac{1}{2\pi i}\frac{\beta^{\prime d/2}(-\delta\lambda_1)^{1-d/2}}{(1+\beta^\prime)^{d/2}}
e^{w\left[\tau p -\frac{\beta^\prime}{2(\beta^\prime+1)}\right]}.
\end{equation}

This integral can be simplified by modifying the integral contour $[1-i\infty, 1+i\infty]$
into a composite of two straight lines of $[-i\infty, -i\epsilon]$
and $[+i\epsilon, +i\infty]$ and a semicircle with an infinitesimally small radius $\epsilon$ to avoid the singular point at the origin ($v=0$). By changing the variable to $y$
as $v=iy$, the integration along the two straight lines becomes a real-valued integral
and the contribution from the semicircle contour can be also done, using
the polar coordinate representation. Summing up these contributions,
one can finally come up with a single real-valued integral expression as in Eq.\ (A16) of
Ref.\ \cite{jslee2}. Then, it is possible to integrate even the tail part of the PDF numerically with very high precision.

In this paper, we just present the results only in
the two scaling regimes of
$(p+1)\tau^{1/2} \ll -1$ and $|p+1|\tau^{1/2} \ll 1$.
In addition, we restrict ourselves to the case of $1\le d <4$ for simplicity.
For $(p+1)\tau^{1/2} \ll -1$, we find
\begin{eqnarray}
P_\tau(w\tau p) &=& \frac{2\beta^{\prime d/2}(w\tau |p+1|)^{\frac{d-2}{2}}}{\pi(1+\beta^\prime)^{d/2}}\Gamma(2-d/2) \nonumber \\
& \times & \frac{\sin\frac{d\pi}{2}}{2-d} \ e^{w\left[\tau p-\frac{\beta^\prime}{2(\beta^\prime+1)}\right]},
\end{eqnarray}
where the term $\sin(d\pi/2)/(2-d)$ goes to $\pi/2$ as $d\rightarrow 2$.
The $\tau$-dependent LDF is given as
\begin{equation}
h_\tau(p)= h_l(p) -\frac{d-2}{2w\tau}\ln \tau -\frac{r_{l}(p)}{w\tau},
\label{eq:tLDFl}
\end{equation}
where $h_l(p)=-p$ and $r_l(p)$ comes from the ${\cal O}(1)$ terms.

For $|p+1|\tau^{1/2} \ll 1$ (a narrow scaling region between the center and the left wing), we find
\begin{eqnarray}
P_\tau(w\tau p) &=& \frac{2{\beta^\prime}^{d/2}(w\tau)^{\frac{d-2}{4}}}{\pi(1+\beta^\prime)^{d/2}}\Gamma(3/2-d/4)\nonumber\\
& \times & \frac{\cos\frac{d\pi}{4}}{2-d}\ e^{w\left[\tau p-\frac{\beta^\prime}{2(\beta^\prime+1)}\right]},
\end{eqnarray}
where the term $\cos(d\pi/4)/(2-d)$ goes to $\pi/4$ as $d\rightarrow 2$.
The $\tau$-dependent LDF is
\begin{equation}
h_\tau(p)= h_l(p) -\frac{d-2}{4w\tau}\ln\tau -\frac{r_{l,c}(p)}{w\tau},
\label{eq:tLDFlc}
\end{equation}
where $h_l(p)$ is the same as that in Eq.\ (\ref{eq:tLDFl}) and $r_{l,c}(p)$ also comes from the ${\cal O}(1)$ terms.

\subsubsection{\label{subsubsec:level323}The right wing of the PDF}

In the right side of the central region $(p\gtrsim 2\beta^\prime +1)$,
the saddle point approaches the singularity at $\lambda=\lambda_2$ $(=-\beta^\prime)$.
In this case, we have an additional complication
due to the essential singularity in the prefactor. Let us write $\delta\lambda_2=\lambda^*-\lambda_2$ $(>0)$. The saddle-point equation (\ref{eq:elp}) is expanded in terms of $\delta \lambda_2$ as
\begin{equation}
-2\delta\lambda_2-(p-1-2\beta^\prime)+\frac{d+2w\beta^\prime
}{2w\tau \delta\lambda_2}+\frac{\beta^{\prime 2}}{2\tau\delta\lambda_2^2}\approx 0.
\label{eq:elpbi}
\end{equation}
Compared to Eq.\ (\ref{eq:elpmi}), it contains a more divergent (fourth) term for finite $\beta^\prime$ and leads to different scaling behavior of $\delta\lambda_2$.
(The case for $\beta^\prime\ll (\delta\lambda_2)^{1/2}$ will be discussed in the next subsection).

For $[p-(2\beta^\prime+1)]\tau^{1/3}\gg 1$, we get
\begin{equation}
\delta\lambda_2\approx\frac{\beta^\prime}{\sqrt{2[p-(2\beta^\prime+1)]}}\tau^{-1/2},
\label{eq:asympt_right}
\end{equation}
which determines the PDF in the most region of $p>2\beta^\prime +1$.

For $|p-(2\beta^\prime+1)|\tau^{1/3}\ll 1$, we get
\begin{equation}
\delta\lambda_2\approx \left(\frac{\beta^{\prime 2}}{4}\right)^{1/3}\tau^{-1/3},
\label{eq:asympt_right_0}
\end{equation}
which determines the PDF in a narrow region around $p=2\beta^\prime +1$ between the center and the right wing of the PDF.

Similar to the left wing, by expanding $H(\lambda,p)$ in powers of $\delta\lambda_2$
and using a new variable $v=1+(\lambda-\lambda^*)/\delta\lambda_2$, we find
\begin{eqnarray}
P_\tau(w\tau p) & = & C_2\int_{1-i\infty}^{1+i\infty}\mathrm{d}v \frac{e^{w\tau\left[(p-2\beta^\prime-1)\delta\lambda_2v+\delta\lambda_2^2 v^2 \right]}}{v^{d/2}} \nonumber \\
& \times & e^{w\frac{\beta^{\prime 2}}{2\delta\lambda_2v}},
\label{eq:PDF_right}
\end{eqnarray}
where
\begin{eqnarray}
C_2 & = & \frac{1}{2\pi i}\frac{\beta^{\prime d/2}\delta\lambda_2^{1-d/2}}{(1+\beta^\prime)^{d/2}} e^{w\tau [\beta^\prime(1+\beta^\prime)-\beta^\prime p]} \nonumber \\
& \times & e^{-w\left(\frac{5}{2}\beta^\prime+2\beta^{\prime 2}\right)}.
\end{eqnarray}
Note that the integrand in Eq.\ (\ref{eq:PDF_right}) has an exponentially diverging
term near $v=0$, which makes useless the previous contour deformation in the left wing
in this case. This makes difficult to evaluate the integral systematically.
Thus, we try to employ again the saddle point method to evaluate this integral
up to ${\cal O}(\tau^{-1})$.

First, for $[p-(2\beta^\prime+1)]\tau^{1/3}\gg 1$, we plug $\delta\lambda_2$ given in Eq.\ (\ref{eq:asympt_right}) into the integrand of Eq.\ (\ref{eq:PDF_right}). Then, the integral without the multiplicative constant $C_2$ can be written as
\begin{equation}
I_a=\int_{1-i\infty}^{1+i\infty}\mathrm{d}v\frac{e^{w\beta^\prime\sqrt{\frac{\tau(p-2\beta^\prime-1)}{2}}\left(v+\frac{1}{v}\right) +\frac{w\beta^{\prime 2}}{2(p-2\beta^\prime-1)} v^2}}{v^{d/2}}.
\end{equation}
Since $\sqrt{\tau(p-2\beta^\prime-1)}\gg\tau^{1/3}$, one can use the saddle-point approximation for the integral.
The saddle point is approximately determined from $\frac{\mathrm{d}}{\mathrm{d}v}(v+1/v)=0$
(the second term in the exponent is much smaller than the first one), yielding $v^*\approx 1$. This saddle point is far from the singularity at $v=0$, so the conventional saddle point integral is sufficient. The curvature proportional to $\frac{\mathrm{d}^2}{\mathrm{d}^2 v}(v+1/v)|_{v=1}=2$ is positive, so the steepest descent path is coincident with the original contour.
As a result, we find
\begin{equation}
I_a= \frac{i 2^{1/4}\sqrt{\pi} e^{w\beta^\prime\sqrt{2\tau(p-2\beta^\prime-1)}+\frac{w{\beta^\prime}^2}{2(p-2\beta^\prime-1)}}}{(w\beta^\prime)^{1/2}[\tau(p-2\beta^\prime-1)]^{1/4}}.
\label{eq:righti1}
\end{equation}
Multiplying it by $C_2$, we get
\begin{eqnarray}
P_\tau(w\tau p)
& = & \sqrt{\frac{\beta^\prime}{\pi w}}\frac{2^{\frac{d-5}{4}}e^{-w\beta^\prime\left[\tau(p-\beta^\prime-1)-\sqrt{2\tau(p-2\beta^\prime-1)}\right]}}{(1+\beta^\prime)^{d/2}\left[\tau(p-2\beta^\prime-1)\right]^{\frac{3-d}{4}}} \nonumber \\
& \times &  e^{-\frac{w\beta^\prime}{2}\left[(5+4\beta^\prime)-\frac{\beta^\prime}{p-2\beta^\prime-1}\right]}.
\end{eqnarray}
The $\tau$-dependent LDF is given as
\begin{equation}
h_\tau(p)= h_r(p) -\frac{\beta^\prime\sqrt{2(p-2\beta^\prime-1)}}{\sqrt{\tau}}
-\frac{d-3}{4w\tau}\ln\tau -\frac{r_{r}(p)}{w\tau},
\label{eq:tLDFr}
\end{equation}
where $h_r(p)=\beta^\prime p-\beta^\prime(1+\beta^\prime)$ and $r_r(p)$ comes from the ${\cal O}(1)$ terms.

Next, for $|p-2\beta^\prime-1|\tau^{1/3}\ll 1$,
 $\delta\lambda_2 \sim \tau^{-1/3} $ as in Eq.\ (\ref{eq:asympt_right_0}).
Again, by the power counting, one can easily simplify the integral in Eq.\ (\ref{eq:PDF_right}) without $C_2$ as
\begin{equation}
I_b=\int_{1-i\infty}^{1+i\infty}\mathrm{d}v \frac{e^{w\tau \delta\lambda_2^2 v^2
+\frac{w\beta^{\prime 2}}{2\delta\lambda_2v}}}{v^{d/2}}.
\label{eq:asympt_right_1}
\end{equation}
A nuisance comes in when we calculate the LDF exactly up to $\mathcal{O}(\tau^{-1})$ (or $\ln P_\tau$ up to $\mathcal{O}(1)$) because higher-order expansions are needed for
$\delta\lambda_2$ in a very narrow region like $|p-2\beta^\prime-1|\sim \tau^{-\alpha}$ with $1/3\le\alpha\le 2/3$. In fact, we need to divide this region into infinitely many
intervals in order to calculate the finite-time correction to the LDF exactly up to
$\mathcal{O}(\tau^{-1})$. This can be done with a straightforward calculation in principle, but requires a lengthy one, involving high-order calculations of $\delta \lambda_2$ from Eq.\ (\ref{eq:elpbi}).

In this paper, we consider only the simplest case of $|p-2\beta^\prime-1| \tau^{2/3}\ll 1$.
Then, both terms in the exponent of Eq.\ (\ref{eq:asympt_right_1})
scale as $\sim \tau^{1/3}$ and the saddle point is determined by $\frac{\mathrm{d}}{\mathrm{d}v}(v^2+2/v)=0$, which gives $v^*\approx 1$. The curvature is also positive, so the steepest path
is again coincident with the original. As a result, we get
\begin{equation}
I_b = i\sqrt{\frac{2^{4/3}\pi}{3w\tau^{1/3}\beta^{\prime 4/3}}}
e^{3w \tau^{1/3}\left(\frac{\beta^\prime}{2}\right)^{4/3}}.
\label{eq:righti2}
\end{equation}
Multiplying it by $C_2$, we obtain
\begin{eqnarray}
P_\tau(w\tau p)&=&\frac{2^{\frac{d-3}{3}}\tau^{\frac{d-3}{6}}\beta^{\prime d/6}}{\sqrt{3\pi w}(1+\beta^\prime)^{d/2}}e^{-w{\beta^\prime}\tau(p-\beta^\prime-1)} \nonumber \\
& \times & e^{3w \tau^{1/3}\left(\frac{\beta^\prime}{2}\right)^{4/3}-\frac{w\beta^\prime}{2}(5+4\beta^\prime)}.
\end{eqnarray}
The $\tau$-dependent LDF is
\begin{equation}
h_\tau(p)= h_r(p)-\frac{3}{\tau^{2/3}}\left(\frac{\beta^\prime}{2}\right)^{4/3}
-\frac{d-3}{6w\tau}\ln\tau -\frac{r_{r,c}(p)}{w\tau},
\label{eq:tLDFrc}
\end{equation}
where $h_r(p)$ is the same as that in Eq.\ (\ref{eq:tLDFr}) and $r_{r,c}(p)$ also comes from the ${\cal O}(1)$ terms.
An extension to higher dimensions $(d\geq 4)$ is straightforward.

\subsubsection{\label{subsubsec:level324}FT violations}

We examine the detailed FT for the heat by varying the initial temperature $\beta^\prime$. We present $f_{\tau}(p)$ defined in Eq.\ (\ref{eq:fq})
such that $f_{\tau}(p)= - h_\tau (p)+h_\tau(-p)$. All the results in this subsection
are summarized into
\begin{widetext}
\begin{equation}
f_\tau(p)=\left\{\begin{array}{ll}p+\frac{1}{\tau}\left[\frac{2\beta^{\prime 2}p}{(2\beta^\prime+1-p)(2\beta^\prime+1+p)}+\frac{d}{2w}\ln\frac{(1-p)(2\beta^\prime+1+p)}{(1+p)(2\beta^\prime+1-p)} \right] & \text{for}\ 0\le p < 1 \\
p-\frac{(1-p)^2}{4}-\frac{d}{4w\tau}\ln\tau+\frac{r_c(p)-r_{l,c}(-p)}{w\tau} & \text{for}\ |p-1|\ll \tau^{-1/2} \\
p-\frac{(1-p)^2}{4}-\frac{d-1}{2w\tau}\ln\tau+\frac{r_c(p)-r_l(-p)}{w\tau} & \text{for}\ 1< p< 2\beta^\prime+1 \\
(1-\beta^\prime)p+\beta^\prime(1+\beta^\prime)+\frac{3}{\tau^{2/3}}\left(\frac{\beta^\prime}{2}\right)^{\frac{4}{3}}-\frac{2d-3}{6w\tau}\ln\tau+\frac{r_{r,c}(p)-r_l(-p)}{w\tau} & \text{for}\ |p-2\beta^\prime-1|\ll \tau^{-2/3} \\
(1-\beta^\prime)p+\beta^\prime(1+\beta^\prime)+\beta^\prime\sqrt{\frac{2(p-2\beta^\prime-1)}{\tau}}-\frac{d-1}{4w\tau}\ln\tau+\frac{r_r(p)-r_l(-p)}{w\tau} & \text{for}\ p-2\beta^\prime-1\gg\tau^{-1/3}, \end{array}\right.
\label{eq:fqsp}
\end{equation}
\end{widetext}
which converge to Eq.\ (\ref{eq:ssft}) for large $\tau$ with various finite-time
corrections.

\begin{figure}[ht!]
\begin{center}
\includegraphics[width=\columnwidth]{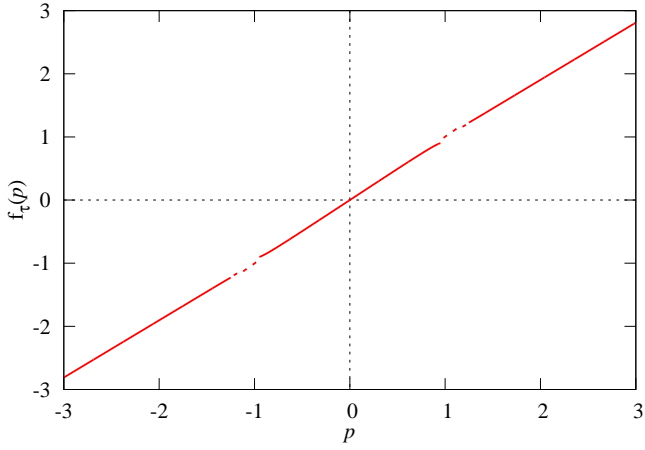}\\%
\end{center}
\caption{\label{fig:fbetap1p10t20nscco}(Color online) $f_{\tau}(p)=\frac{1}{w\tau}\ln \frac{P_{\tau}(w\tau p)}{P_{\tau}(-w\tau p)}$ is drawn for $\beta^\prime=0.1$, $\tau=20$, and $d=3$.}
\end{figure}
\begin{figure}[ht!]
\begin{center}
\includegraphics[width=\columnwidth]{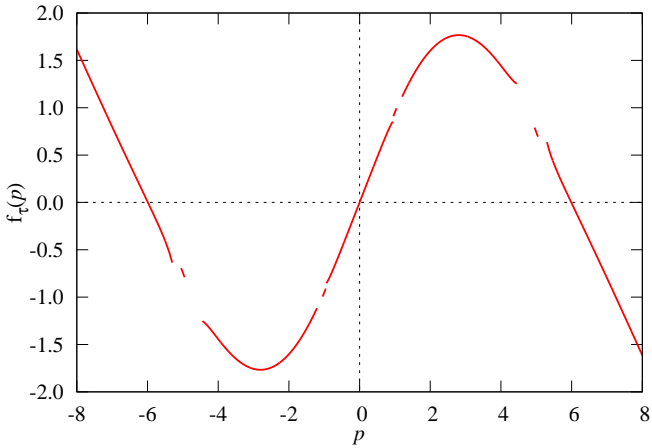}\\%
\end{center}
\caption{\label{fig:fbetap2t20fgccoa}(Color online) $f_{\tau}(p)$ is drawn for $\beta^\prime=2$, $\tau=20$, and $d=3$.}
\end{figure}

Inside of the region I ($|p|<1$), the detailed FT is violated for finite $\tau$
by the amount of ${\cal O}(1/\tau)$, and
the FT is restored $(f_\infty(p)=p)$ in the infinite-$\tau$ limit.
In all other regions, the FT is violated even in the infinite-$\tau$ limit.
We present the figures for $f_{\tau}(p)$, Fig.\ \ref{fig:fbetap1p10t20nscco}
for $\beta^\prime=0.1$ and Fig.\ \ref{fig:fbetap2t20fgccoa} for $\beta^\prime=2$.
They show similar trends to $f_{\infty}(p)$ as in Fig.\ \ref{fig:fqbetap01}.
The FT holds approximately well only in the central region (I).
%

\subsection{\label{subsec:level33}FT in the $\beta^\prime\to 0$ limit}

In this subsection, we establish the transient detailed FT for the heat in general, from
the standard stochastic thermodynamics where the time-integrated quantities
are defined at the level of dynamic trajectories~\cite{sekimoto,seifert,esposito}.

A trajectory starting from $t=0$ to $\tau$, is denoted by $\mathbf{q}(t) \equiv \{q_t; t \in [0,\tau]\}$ with a set of state variables
$q_t$. The probability to find  a trajectory $\mathbf{q}$ in a given dynamic process
can be written as
\begin{equation}
P (\mathbf{q})
= P_0 (q_0) \Pi (\mathbf{q} | q_0),
\label{eq:1}
\end{equation}
where $P_0 (q_0)$ is the probability distribution of the initial state $q_0$
and $\Pi (\mathbf{q} | q_0)$ is the conditional probability for
the trajectory $\mathbf{q}$ starting from $q_0$.

We also define the time-reverse trajectory $\mathbf{q}^\dagger$ with
$\mathbf{q}^\dagger (t)=\epsilon{\mathbf{q}}(\tau-t)$ with $\epsilon{\mathbf{q}}$
represents the mirrored trajectory with a parity $\epsilon$ for each state variable~\cite{spinney,hklee}. This trajectory starts at the mirrored
state of the final state of the original trajectory: $q^\dagger_0=\epsilon {q}_\tau$.
The trajectory probability for $\mathbf{q}^\dagger$ is similarly written as
\begin{equation}
P (\mathbf{q}^\dagger)
= P_0 (q^\dagger_0)\ \Pi (\mathbf{q}^\dagger | q^\dagger_0).
\label{eq:2}
\end{equation}

It is well known~\cite{seifert,esposito} that the heat production for a given trajectory $\mathbf{q}$ is
identical to the logarithm of the ratio of two conditional probabilities as
\begin{equation}
\beta Q_\tau [\mathbf{q}]= \ln \frac{\Pi (\mathbf{q} | q_0)}{\Pi (\mathbf{q}^\dagger | q^\dagger_0)},
\label{eq:heat}
\end{equation}
where $\beta$ is the inverse temperature of the heat bath.

By choosing various initial ensembles for the original and the time-reverse processes
[$P_0(q_0)$ and $P_0 (q^\dagger_0)$], one can derive FTs for different thermodynamic
quantities. For example, when one chooses the initial distribution of the time-reverse process as the final distribution  of the original process [$P_0 (q^\dagger_0)=P_\tau(q_\tau)$], then the total entropy production summing the system's Shannon entropy change and heat production becomes simply a logarithm of the ratio of two trajectory probabilities such that $\Delta S_{\rm total} = \ln [ P (\mathbf{q})/P (\mathbf{q}^\dagger)]$. Since the $\Delta S_{\rm total}$ is written as the logarithm of the two normalized probability distributions (a typical property of the relative entropy), the integral FT should hold for $\Delta S_{\rm total}$ for any finite $\tau$ and any initial ensemble with $P_0(q_0)$~\cite{seifert,esposito}. For the transient detailed FT, we need the so-called involution condition, which requires the steady-state initial ensemble.

It is also well known that the choice of the equilibrium Boltzmann
ensembles as the initial ensembles for both the original and time-reverse processes yields the transient integral and detailed FTs for the work, where the involution condition is automatically satisfied with this choice.

We can also derive the FT for the heat  in a similar manner by choosing the uniform (state-independent) distributions as the initial distributions for both processes. Then,
the logarithm of the ratio of trajectory probabilities is simply the heat production
as in Eq.\ (\ref{eq:heat}), due to the cancelation of $P_0(q_0)$ and $P_0(q^\dagger_0)$.
Since these initial distributions are obviously involutary to each other, not only the
integral but also the detailed FT should hold for any finite $\tau$.
Even though the uniform distribution cannot be realized in the infinite-state space,
one may consider it as the infinite-temperature ($\beta^\prime=0$) limit of the Boltzmann distribution.

In the $\tau=\infty$ limit, we have already shown that the detailed FT is satisfied by
taking the $\beta^\prime=0$ limit as in Eq.\ (\ref{eq:ssft}). However, the finite-time
corrections in Eq.\ (\ref{eq:fqsp})
seem to indicate that the  $\beta^\prime=0$ limit does not restore the FT for finite $\tau$.
This suggests the non-commutativity between the $\tau\rightarrow\infty$ limit and
the $\beta^\prime\rightarrow 0$ limit, which indeed turns out to be true.

All the complications come from the calculation of the PDF in the right wing.
The saddle point equation in Eq.\ (\ref{eq:elpbi}) has the $\beta^\prime$-dependent
divergent (fourth) term. In the case that $\beta^\prime$ is small and approaches zero for large $\tau$ such that $\beta^\prime\ll (\delta\lambda_2)^{1/2}$, this fourth
term can be ignored with respect to the third term. Then, all the subsequent calculations are very similar to those for the left wing of the PDF. The results are summarized below.

For $(p-1)\tau^{1/2}\gg 1$, $\delta\lambda_2\approx \frac{d}{2w (p-1)}\tau^{-1}$ and
\begin{equation}
h_\tau(p)= h_r(p) -\frac{d-2}{2w\tau}\ln\tau -\frac{\tilde{r}_{r}(p)}{w\tau},
\label{eq:tLDFlb}
\end{equation}
with $h_r(p)=\beta^\prime p-\beta^\prime(1+\beta^\prime)$. Note that $\tilde{r}_r(p)=r_l(-p)$ in Eq.\ (\ref{eq:tLDFl}) for $\beta^\prime \ll (\delta\lambda_2)^{1/2}$.

For $|p-1|\tau^{1/2} \ll 1$ , $\delta\lambda_2\approx \frac{1}{2}\sqrt{d/w}\ \tau^{-1/2}$ and
\begin{equation}
h_\tau(p)= h_r(p) -\frac{d-2}{4w\tau}\ln\tau -\frac{\tilde{r}_{r,c}(p)}{w\tau},
\label{eq:tLDFlcb}
\end{equation}
where $\tilde{r}_{r,c}$ is a $p$-independent constant and equal to $r_{l,c}$ in Eq.\ (\ref{eq:tLDFlc}) for $\beta^\prime \ll (\delta\lambda_2)^{1/2}$. In both cases, $\delta\lambda_2$ decays
 with $\tau$, so does $\beta^\prime$.

In the calculation of $f_\tau=-h_\tau(p)+h_\tau(-p)$, nice cancelations occur
between the finite-time correction terms  up to ${\cal O}(\tau^{-1})$, and the FT
is fully restored in the region I ($|p|<1$). However, in the other regions,
we still have an extra term such as in $f_\tau(p)=-h_r(p)+h_l(-p)\approx p+\beta^\prime (1-p)$. This extra term may still be bigger than ${\cal O}(\tau^{-1})$ with the $\tau$-dependent $\beta^\prime$, satisfying the condition $\beta^\prime\ll (\delta\lambda_2)^{1/2}$. Therefore, the full FT for finite $\tau$ should be restored only in the $\beta^\prime\rightarrow 0$ limit before any large-$\tau$ limit is taken.

\section{\label{sec:level4}LDF and FT for work}

The generating function $g_W(\lambda)$ for the work is Gaussian in $\lambda$
without any singularity. Thus, its Fourier integration in Eq.\ (\ref{eq:pdf}) can be evaluated exactly. Using a scaled variable for the work $\tilde{p}\equiv W/(w\tau)$,
we find an exact PDF including all the exponentially decaying terms as
\begin{equation}
P_{\tau}(w\tau\tilde{p}) = \frac{e^{-\frac{w[\tau(\tilde{p}-1)+1-\alpha]^2}{2\left[2\tau-3+1/\beta^\prime+\alpha b(\alpha) \right]}}}{\left\{2\pi w\left[2\tau-3+1/\beta^\prime+\alpha b(\alpha) \right]\right\}^{1/2}},
\end{equation}
where the exact generating function in Eq.\ (\ref{eq:origg2}) was integrated
with $\alpha=e^{-\tau}$ and $b(\alpha)=2(2-1/\beta^\prime)-\alpha(1-1/\beta^\prime)$.

Then, the $\tau$-dependent LDF is simply given by
$h_\tau(\tilde{p})=-(w\tau)^{-1} \ln P_{\tau}(w\tau\tilde{p})$ as
\begin{equation}
h_\tau(\tilde{p}) = h_W(\tilde{p}) +\frac{\ln \tau }{2w\tau} -\frac{r_W(\tilde{p})}{w\tau},
\label{eq:tLDFW}
\end{equation}
where $h_W(\tilde{p})=(\tilde{p}-1)^2/4$ and $r_W(\tilde{p})$ comes from the ${\cal O}(1)$ terms.
The FT-examining function $f_\tau(\tilde{p})$ becomes
\begin{equation}
f_\tau(\tilde{p}) =\frac{2 \tilde{p}(\tau-1+\alpha)}{2\tau-3+1/\beta^\prime +\alpha b(\alpha)}.
\end{equation}
At $\beta^\prime=1$, $b(\alpha)=2$ and $f_\tau(\tilde{p})=\tilde{p}$ exactly
for any $\tau$. Thus, the transient detailed FT holds exactly at $\beta^\prime=1$.
For $\beta^\prime\neq 1$, $f_\tau(\tilde{p})\approx \tilde{p}+ {\cal O}(\tau^{-1})$
for large $\tau$, so the detailed FT is satisfied only in the $\tau=\infty$ limit.

\section{\label{sec:level6}Discussion}

The memory of the initial state is shown not to vanish, but to remain in the rare events for the time-accumulated quantities such as heat and work. This novel phenomenon is manifested in the form of PDF particularly in the tail region corresponding to the rare events. The FT for finite $\tau$ depends on the initial ensemble. For example, the work satisfies the transient detailed FT only with an initial equilibrium Boltzmann distribution, while the heat does only with an initial uniform distribution. A common sense suggests that the heat and work satisfy the detailed FTs simultaneously in the long-time limit, since both quantities are equivalent to each other on average. However, it turns out that the FTs for the heat and work in the long-time limit deviate in different ways. In this limit, the FT for work is satisfied with any initial ensemble, while the FT for heat is not valid except for the uniform initial ensemble. This discrepancy originates from the unboundedness of the (potential) energy fluctuations $\Delta U$ in the heat
$Q=W-\Delta U$.

In this paper, we investigate the PDF's for the work and heat generated in the Brownian motion
in a sliding harmonic potential with a general initial ensemble characterized by the Boltzmann distribution with the inverse temperature $\beta^\prime$, generally different from the inverse temperature $\beta$ of the reservoir. The heat PDF is calculated analytically up to
${\cal O}(\tau^{-1})$ with the measuring time $\tau$ and the work PDF is obtained
exactly for any finite $\tau$.

We explicitly show that the transient detailed FT holds for the work only at $\beta^\prime=1$ and for the heat only at $\beta^\prime=0$, as expected. On the other hand,
the detailed FT in the long-time limit holds at any $\beta^\prime$ for the work, but
only at $\beta^\prime=0$ for the heat (one should be careful about the order of the two limiting procedures of $\beta^\prime\rightarrow 0$ and $\tau\rightarrow\infty$).
This is due to the presence of singularities
in the boundary terms for the heat, which represents the persistence of the initial memory.
Physically, it can be argued that the highly energetic particles (high $U$)
in the initial ensemble dominantly contribute to the events of positive large heat production $(Q\gg 1)$ by losing its energy through dissipation~\cite{jslee1}.
This is why the right wing of the heat PDF depends strongly on the initial temperature
$\beta^\prime$, but its left wing depends on it only very weakly.
It is interesting to note that there is no threshold value of $\beta^\prime$ for the dominance of the initial ensemble, in contrast to other cases where a finite critical
threshold is found~\cite{jslee1}.

It may be an interesting task to find a systematic deviation of the FT for time-integrated quantities with an arbitrary initial ensemble, for example, by generalizing the recent study
on the relation of heat fluctuations in Ref.\ \cite{noh}. It is also interesting to apply our method to other solvable nonequilibrium systems, such as a linear diffusion system with a nonconservative force~\cite{kwon,noh-kwon-park} and a motion under a breathing harmonic potential~\cite{kwon1}.

\begin{acknowledgments}
This work was supported by the EDISON program through NRF Grant No.\ 2012M3C1A6035307 (K.K.) and also by the Basic Science Research Program through NRF Grant No.\ 2013R1A1A2011079 (C.K.) and 2013R1A1A2A10009722 (H.P.).
We thank Korea Institute for Advanced Study for providing computing resources (KIAS Center for Advanced Computation Abacus) for this work. H.P.~also thanks the Galileo Galilei Institute for Theoretical Physics for the hospitality and the INFN
for partial support during the completion of this work.
\end{acknowledgments}

\appendix

\section{\label{sec:app1}Generating functions}

The exact formula for the generating function for the heat is given by
\begin{equation}
g_Q(\lambda)
= \left(\frac{\beta^\prime}{\beta^\prime+\lambda}\right)^{d/2}
\frac{1}{C^{d/2}}\ e^{- w G_Q(\lambda,\tau)}
\label{eq:g_Q0}
\end{equation}
with
\begin{eqnarray}
G_Q(\lambda,\tau)
&=& \tau\lambda(1-\lambda) -\frac{3}{2}\lambda +\frac{\lambda^2}{2C}
(1-\lambda)(4-B) \nonumber \\
&+& \alpha \left[2\lambda - \frac{2\lambda^2}{C}
(1-\lambda)(2-B)\right] \nonumber \\
&+& \alpha^2 \left\{-\frac{\lambda}{2} + \frac{\lambda^2}{2C}
\left[ 7-6B - (4-3B) \lambda\right]\right\} \nonumber \\
&-& \alpha^3 \frac{2\lambda^2}{C} (1-B)
+\alpha^4 \frac{\lambda^2}{2C}(1-B)\ ,
\label{eq:origg}
\end{eqnarray}
where
\begin{equation}
B=(\beta^\prime + \lambda)^{-1},\ C=1-\lambda+\alpha^2\lambda(1-B),\ \alpha=e^{-\tau}.
\end{equation}
By setting $\alpha=0$ (in the long-time limit), we get Eq.\ (\ref{eq:g3}) in
Sec.\ \ref{subsec:level22}.
At $\beta^\prime=\beta=1$, we find
\begin{equation}
g_Q(\lambda)=\frac{e^{-w\lambda(1-\lambda)\left\{\tau +\frac{1-\alpha}{\tilde{C}}
\left[2\lambda^2 (1-\alpha) -\frac{1}{2}(3-\alpha)\right]\right\}}}{\tilde{C}^{d/2}},
\end{equation}
with ${\tilde{C}}=1-\lambda^2 +\lambda^2 \alpha^2$. Note that our result
is slightly different from that in Ref.\ \cite{vanzon2}.

In the limit of $\beta^\prime=0$, $g_Q(\lambda)$ vanishes
as $\sim {\beta^\prime}^{d/2}$.
However, note that its amplitude in this limit
\begin{equation}
g_Q(\lambda) \ {\beta^\prime}^{-d/2}=\frac{e^{-w\lambda(1-\lambda)\left[\tau
-2\frac{1-\alpha}{1+\alpha}\right]}}
{\left[\lambda(1-\lambda)(1-\alpha^2)\right]^{d/2}},
\end{equation}
perfectly satisfies the GC symmetry for any $\alpha$ (a finite time).

The generating function for the work is given by
\begin{equation}
g_W(\lambda)=e^{-w G_W(\lambda,\tau)}
\end{equation}
with
\begin{eqnarray}
G_W(\lambda,\tau)
& = &\tau\lambda(1-\lambda) -\lambda
+\frac{\lambda^2}{2}
\left(3-\frac{1}{\beta^\prime}\right) \nonumber \\
&+& \alpha \left[\lambda -\lambda^2\left(2-\frac{1}{\beta^\prime}\right)\right] \nonumber \\
&+& \alpha^2 \frac{\lambda^2}{2} \left(1-\frac{1}{\beta^\prime}\right).
\label{eq:origg2}
\end{eqnarray}
By setting $\alpha=0$, we get Eq.\ (\ref{eq:c}) in
Sec.\ \ref{subsec:level23}.
At $\beta^\prime=\beta=1$, we find
\begin{equation}
g_W(\lambda)=e^{-w\lambda(1-\lambda)(\tau -1+\alpha)},
\end{equation}
which agrees with the result in Ref.\ \cite{vanzon2}
and satisfies perfectly the GC symmetry for any $\alpha$ (a finite time).


\begin{thebibliography}{99}
\bibitem{exp1} $A=\beta(W-\Delta F)$ being taken, the GC symmetry is exact for any finite $\tau$ with the initial Boltzmann ensemble characterized by the inverse temperature $\beta$ of the heat bath.
\bibitem{exp2} The free energy difference is $\Delta F=0$ in this simply driven system.
\bibitem{discret} In the discrete-time representation, $\int_0^{\tau} \mathrm{d}t \mathcal{L}$ is replaced with $-\frac{1}{4D}\sum_{i=1}^{N}\left(\frac{\mathbf{X}_i-\mathbf{X}_{i-1}}{\Delta t}+\tau_r^{-1}\mathbf{X}_i^{(\alpha)}\right)^2-\alpha d\tau_r^{-1} \tau$, where $\mathbf{X}_i^{(\alpha)}=\alpha \mathbf{X}_i+(1-\alpha)\mathbf{X}_{i-1}$ for $\alpha=0$ (prepoint, Ito), $\alpha=1/2$ (midpoint, Stratonovich), and $\alpha=1$ (postpoint). The first term is a contribution from the Lagrangian, while the second term should be kept for a specific representation of $\alpha$. The simplest one is $\alpha=0$, but the path integral is independent of $\alpha$.
\end{thebibliography}
\end{document}